\documentclass{appolb}
\usepackage[a4paper, total={6.5in, 8.5in}]{geometry}
\usepackage{graphicx, color}
\usepackage[sort&compress,numbers,square]{natbib}
\usepackage[colorlinks=true, urlcolor=blue, pdfborder={0 0 0}]{hyperref}
\usepackage{amsmath, amsthm, amssymb}
\usepackage{xcolor}

\usepackage{subcaption}

\newcommand{\fr}{\mathtt{R}}

\newcommand{\frl}{f(\mathtt{R,L_m})}

\begin{document}
\title{In-Stabilities of massive white dwarfs in modified gravity%
\thanks{Presented at 4th Jagiellonian Symposium by Marek Nowakowski}%
}
\author{Ronaldo V. Lobato
\address{Departamento de
    F\'isica, Universidad de los Andes, Bogot\'a, Colombia and
    Department of Physics and Astronomy, Texas A\&M
    University-Commerce, Commerce, TX, USA and ICRANet, P.zza della Repubblica 10, I-65122 Pescara, Italy
  }
\\[3mm]
{Geanderson A. Carvalho 
\address{Departamento de F\'isica, Universidade Tecnol\'ogica Federal
  do Paran\'a, Medianeira, PR, Brazil}}
\\[3mm]
{Neelima G. Kelkar, Marek Nowakowski\address{Departamento de
    F\'isica, Universidad de los Andes, Bogot\'a, Colombia}}
}

\maketitle
\begin{abstract}
  Super-Chandrasekhar white dwarfs are a timely topic in the last years in
the scientific community due to its connection to supernovae type Ia (SN Ia). Some early studies tackled the
possibility of white dwarfs surpassing the Chandrasekhar limit by
means of a magnetic field. More recently modified gravity has been
highlighted as the reason for these stars to surpass the Chandrasekhar
limit and becoming a supernova progenitor. However, in general simple
assumptions are considered for the stellar structure and equation of
state (EoS), which can lead to unreliable conclusions. In this work we
want to be rigorous and consider a realistic EoS to describe the
white dwarfs in general relativity and modified gravity, taking into
account nuclear instabilities that limit the maximum mass.

\end{abstract}

\section{Introduction}

White dwarfs (WD) are stars that can reach densities as high as
$\sim10^{11}\ {\rm g/cm^3}$ in their interiors. Observed magnetic fields
of WDs are also in the order of $\sim 10^9$ G, while the masses are limited
by the so called Chandrasekhar mass limit (maximum stable mass $M_{\rm
  Ch}=1.44~M_\odot$, with $M_\odot$ representing the Solar mass). The
radii of WDs are of order $10^4$ km, which renders a surface gravity,
log$_{10} g$, in the range $8-10$. These extreme properties make WDs a
laboratory of tests for strong gravity regimes, thus motivating their
application to the study of modified gravity theories. In particular,
WDs can help to constrain the parameter space of the new theories.

On the other hand, some peculiar, overluminous type Ia supernovae have
been linked to the possible existence of super-Chandrasekhar white
dwarfs. The origin of type Ia supernovae is understood as the collapse
of either a WD binary or a massive, near Chandrasekhar mass accreting
WD. However, the possible progenitor systems or the mechanism leading to
such massive WDs are highly unknown. Several scenarios were studied: WDs
with rotation
\cite{Boshkayev2011Dec,Becerra2019Jul,Terrero2017Dec,Boshkayev2013Jul,Boshkayev2018Dec,Boshkayev2018Aug},
within modified theories
\cite{Wojnar2021Feb,Olmo2020Sep,Maurya2020Dec,Hansraj2019Jan,Sharif2018Feb,Lobato2020Dec,Lobato2022Jun,Rocha2020May,Carvalho2018May,otoniel/2017,Otoniel2019Jul,Carvalho2017Dec,Sharif2018Jun},
with magnetic and electric fields
\cite{Bera2014Dec,Bera2016Mar,Chamel2013Oct,Coelho2014Sep,Das2013Feb},
temperature \cite{boshkayev/2016, boshkayev/2017, decarvalho/2014a, nunes/2021}, under generalized uncertainty principle, in Einstein-$\Lambda$ gravity \cite{Liu2019Feb}.

\section{Hydrostatic equilibrium}
To model relativistic stars, one needs the general relativity equations
\begin{equation}
  G^{\mu\nu} \equiv R^{\mu\nu} - \frac{1}{2}g^{\mu\nu}R = 8\pi
  T^{\mu\nu}.
\end{equation}

For a perfect fluid energy-momentum tensor and for a static spherical
symmetric spacetime, the Einstein's field equations lead to the
hydrostatic equilibrium equation, the Tolman-Oppenheimer-Volkoff (T.O.V.) equation~\cite{tolman/1939,
  oppenheimer/1939}. This equation reads in natural units
\begin{equation}
\label{TOV-Eq}
  p' = - (\rho + p) \frac{4\pi pr + m/r^2}{(1 - 2m/r)},
\end{equation}
where the prime indicates radial derivative and $m$ is the gravitational mass enclosed within the
surface of radius, $m' = 4\pi\rho r^2$. To solve this system one needs
the previous EoS ($p(\rho)$) and use the boundary conditions
\begin{equation}
m(r)|_{r=0}= 0,
\end{equation}
where $m(r) = 4/3\pi r^3 \rho(0)$ and
\begin{equation}
p(r)|_{r=0}=p_{c}\quad \textrm{and}\quad \rho (r) |_{r=0}=\rho _{c},
\end{equation}
where $p_c$ and $\rho_c$ are the pressure and density at the center of
the star. The numerical integration of Eq.~(\ref{TOV-Eq}) follows the
pressure decrease as one moves away from the center, and it is stopped
when the condition $p(r)|_{r=R}=0$ is reached at the surface of the star $R$. The integration of the profile density
\begin{equation}
M(R) \equiv 4 \pi \int_0^R r^2 \rho(r) dr
\end{equation}
provides the total gravitational mass of the star $M$.  The resulting
M-R relation can be compared to data from astronomical
observations. Once the EoS is provided, the global properties of the
stars can be obtained.

When one considers a modification in gravity theory, the field equations are changed. Generally, a symmetric spacetime/perfect fluid
energy-momentum tensor are still used. In this case, one will have the T.O.V.-like equations for the hydrostatic equilibrium equations that model the relativistic stars.

For the specific theory called $f(\mathtt{R,L_m})$ gravity, that we
have considered before \cite{Lobato2022Jun}, where $f(\mathtt{R,L_m})=\mathtt{R}/2+\mathtt{L_m}+\sigma \mathtt{R L_m}$, the equations are:

\begin{subequations}\label{tov-like}
  \begin{eqnarray}
    \alpha'(p+\rho) + 2z = 0,
    \end{eqnarray}
    \begin{eqnarray}
      p' - z = 0,
    \end{eqnarray}
    \begin{eqnarray}
      &\Bigg[\bigg(2 r^{2} \rho e^{\beta} + \big(2  \fr r^{2} \rho e^{\beta} + 3 r^{2} z
             \alpha'+ 6  p r \beta' + 2  \big(2  \fr p r^{2} + 3  p\big) e^{\beta} - 6
        p\big) \sigma \nonumber \\
        &- \left({\left(\fr - 3  p\right)} r^{2} + 3\right) e^{\beta} - 3  r \beta' +
        3\bigg) e^{-\beta}\Bigg]({3r^{2}})^{-1} = 0,
    \end{eqnarray}
    \begin{eqnarray}
     & \Bigg[\bigg(r^{2} \rho e^{\beta} + (\fr r^{2} \rho e^{\beta} + 3 \, r^{2} z \beta' + 6
       \, p r \alpha' - 6 \, r^{2} z' - (\fr p r^{2} + 6 \, p)
       e^{\beta} + 6 \, p) \sigma \nonumber \\ & + {\left(\fr r^{2} + 3\right)} e^{\beta} - 3 \, r
       \alpha' - 3\bigg) e^{-\beta}\Bigg]({3 \, r^{2}})^{-1} = 0.
\end{eqnarray}
\end{subequations}
where, $\alpha$ and $\beta$ are the metric potentials depending on the
radial coordinate $r$ and $z$ is an auxiliary variable, $z=p'$. For complete details, see
Refs.~\cite{carvalho/2020, lobato/2021}. Once the EoS is defined,
the global properties, such as mass and radius, can be found from \eqref{tov-like}.

\section{Stability criteria for critical mass}
The critical mass of white dwarfs is known from a long time ago, when
Stoner \cite{stoner/1929} considered special relativity to
describe Fermi-Dirac statistics stars, the mass was established as,
\begin{equation}
  M_{\rm crit} \approx K\frac{M^3_{\rm P}}{\mu^2m^2_n},
\end{equation}
where $M_{\rm P}$ is the Planck mass, $m_n$ is the neutron mass and
$\mu$ the average molecular weight $A/Z$. The constant $K$ was
determined as $K=3.72$, later in the Chandrasekhar \cite{chandrasekhar/1931, chandrasekhar/1935},
Landau \cite{landau/1938} and Gamow \cite{gamow/1939} works, the value went to $K=3.09$ using
Lane-Emden equations. To reach this value, the simplest EoS was used, it
considers a model of non-interacting relativistic Fermi gas of electrons, although the EoS can
describe very well WDs, there were improvements such as the one by
Hamada-Salpeter (HS), which accounts for corrections due electrostatic
energy, Thomas-Fermi deviations, exchange energy and spin-spin
interactions \cite{hamada/1961, salpeter/1961}. However, only
electrostatic corrections were found to be non-negligible. The
Chandrasekhar EoS has a dependence on $\mu$, and HS besides a dependence on $\mu$ one
changes the dependence on the nuclear composition of the star, for the
have on $Z$ also, which slightly decreases the Chandrasekhar
limit as one can see in Fig. \ref{mrall}, where we have the mass-radius
 for white dwarfs considering the HS EoS for different star's composition.

The electron pressure in HS EoS is lowered by the electrostatic
attraction among the electrons and ions. Further and new developments were considered when heavy elements are
important, in the Thomas-Fermi \cite{haensel/2007} and
Feynman-Metropolis-Teller models \cite{feynman/1949}. The hole of
electron-ion interaction started to be considered in these models in
more ways,
i.e., inclusion of corrections of nuclear thresholds
such as inverse $\beta$-decay and pycnonuclear reactions
\cite{adam/1985, adam/1986},
leading to study of the low mass neutron stars, that could be
\cite{nomoto/1991, janka/2012}, i.e., massive WDs near the
Chandrasekhar limit \cite{brachwitz/2000}.
generated by massive white dwarfs made of oxygen-neon-magnesium.

\subsection{Gravitational instability}
When considering perturbations around the static equilibrium, one can start from the perturbed Euler equations,
\begin{equation}\label{euler}
\Delta\left(\frac{dv^i}{dt}+\frac{1}{\rho}\nabla_i p + \nabla_i \Phi\right)=0,
\end{equation}
where $\Phi$ is gravitational potential. Note that the unperturbed equations with $\vec{v}=0$ (static configuration) lead to the equation of hydrostatic equilibrium. One can assume $\Delta \vec{v}={\rm d}\xi/{\rm d}t$, being $\xi$ perturbations of the form $\xi=\xi(\vec{x},t)$, and commutation relations between $\Delta$ and derivatives. From \eqref{euler}, one arrive at,
\begin{equation}
    \rho \frac{{\rm d}^2\xi^i}{{\rm d}t^2}-\frac{\Delta \rho}{\rho}\nabla_i p+ \nabla_i\Delta P+ \rho \nabla_i\Delta\Phi=0,
\end{equation}
which gives the dynamical equation for the perturbations. If the perturbations are now restricted to follow $\xi(\vec{x},t)=\xi(\vec{x}){\rm e}^{i\omega t}$, the problem becomes a Sturm-Liouville eigenvalue equation for $\omega^2$. The eigenvalues form an infinite and discrete sequence $\omega_0^2<\omega_1^2<\omega_2^2$, where $\omega_0$ is the oscillation frequency of the fundamental mode. The eigenvalues are expected to be real, so $\omega^2<0$ correspond to unstable oscillation modes. One interesting case is when the oscillation frequency corresponds to radial perturbations, $\xi(\vec{x})=\xi(r)$. In this situation, one can show that the oscillation frequency become
\begin{equation}\label{wprop}
    \omega^2\propto 3\bar{\Gamma}_1-4,
\end{equation}
where $\bar{\Gamma}_1$ is the pressure-average adiabatic index, i.e.,
\begin{equation}
    \bar{\Gamma}_1\equiv \frac{\int_0^R\Gamma_1 pr^2{\rm d}r}{\int_0^R pr^2{\rm d}r},
\end{equation}
with $\Gamma_1$ the adiabatic index governing the perturbations, i.e., $\Delta P/P=\Gamma_1 \Delta\rho/\rho$.

When a one-parameter sequence of equilibrium stars is constructed with an
EoS considering different central densities, the critical point $dE/d\rho_c = 0$ gives also the onset of instability. In particular, using the variational principle and an adiabatic EoS $p=K\rho^\Gamma$, it is possible to show that \cite{shapiro/2008}
  \begin{equation}\label{dmdrhos}
    \frac{\partial M}{\partial \rho_c} \propto \Gamma - \frac{4}{3}.
    \end{equation}
From equation \eqref{wprop} $\Gamma<4/3$ gives a negative $\omega^2$, so leading to unstable configurations under radial perturbations. On the other hand, if $\Gamma>4/3$, positive values of $\omega^2$ are achieved, now giving a region of stable configurations. This correlated to equation \eqref{dmdrhos} translates into
\begin{equation}
    \frac{\partial M}{\partial \rho_c}>0, \quad {\rm for~ stable~ equilibrium~ configurations}
\end{equation}
and
\begin{equation}
    \frac{\partial M}{\partial \rho_c}<0, \quad {\rm for~ unstable~ equilibrium~ configurations}.
\end{equation}
So, if we have only one critical point in a $M$-$\rho_c$ equilibrium
sequence, it marks the onset of stability under radial oscillations, defining a maximum mass allowed due to gravitation. In general, the
works that have studied white dwarfs in modified gravity applied only this
gravitational stability criteria, and, in addition, they have used a simplistic Chandrasekhar EoS. When improvements in
the EoS as seen previously are considered,
the maximum mass decreases. Moreover, when
considering the onset of nuclear instabilities, they are often reached before the gravitational instability, which limits even the maximum mass in GR \cite{chamel/2013}. That is also
important for modified gravity, i.e., the onset of nuclear instability should
be taken into account since it will turn on before the
gravitational one.

\subsection{Nuclear instabilities}
\subsubsection{Inverse $\beta$-decay}
The first cutoff to the Chandrasekhar equation of state due to nuclear
reactions comes from
the effects of inverse $\beta$-decay, which reduces the maximum mass
$M$ of the white dwarfs \cite{ostriker/1968a}. That occurs through the
electron capture when the electron Fermi energy is larger than the
mass in the initial and final state. As the star goes to higher
density, the matter suffers compression and the electrons combine with
nuclei, generating another nucleus and a neutrino \cite{chamel/2015},
\begin{equation}
^{A}_{Z}X + e^{-} \rightarrow ^{A}_{Z-1}Y + \nu_{e}.
\end{equation}
The electron capture leads to a global instability of the star, that
can induce a core-collapse of the white dwarf, to undergo collapse
depends on the relation between the electron capture and the
pycnonuclear reactions. The instability of a pure $^{12}$C star,
considering the general relativistic effects, has been
calculated \cite{canal/1976}, leading to a maximum mass of $M\approx
1.366M_{\odot}$. It was computed that the maximum Fermi energy of the
electrons is 12.15 MeV, and mixed WD of $^{12}$C/$^{16}$O, the
configuration becomes unstable when the $^{16}$O concentration exceeds
0.06, leading to a maximum mass of $M\approx
1.365M_{\odot}$.

For the reaction occurs one needs that the Gibbs energy per nucleon
should be higher that of the Gibbs energy of the nucleon produced, so
the condition below should be satisfied
\begin{equation}
g(p,
A, Z) \geq g(p,
A, Z-1).
\end{equation}
For a detailed discussion about neutronization, see Sec. V. of Ref. \cite{chamel/2015}.

\subsubsection{Pycnonuclear reaction}
A second cutoff in maximum mass of the WDs EoS is due to the pycnonuclear
reactions. The screen of Coulomb potential due to the electrons in the
lattice that composes the white dwarfs make the potential barrier of
the nuclei decreases, easier to cross. For higher densities as in
the core of white dwarfs the oscillations of the nuclei in the lattice
produce pycnonuclear reactions, these reactions can be written as
\cite{chamel/2013}
\begin{equation}
^{A}_{Z}X + ^{A}_{Z}X \rightarrow ^{2A}_{2Z}Y.
\end{equation}
As the elements fuses, the threshold for the pressure reduces as the
resulting nucleus has a lower electron capture threshold
\cite{chamel/2014}. As pointed out by Ref. \cite{chamel/2013} the
rates which the pycnonuclear reactions happens is very uncertain. The
threshold is defined by the electron capture that is lower for the
$^{2A}_{2Z}Y$ than for $^{A}_{Z}X$.

\section{Results and discussions}

Considering the Hamada-Salpeter EoS \cite{salpeter/1961} for $^4$He, $^{12}$C, $^{16}$O,
$^{20}$Ne, $^{24}$Mg,  $^{32}$S, $^{56}$Fe and using the mass density threshold for
electron capture of Ref. \cite{chamel/2015}, see Table \ref{limits}, we have constructed stellar sequences of equilibrium
considering general relativity and $f(\mathtt{R,L_m})$ theory of gravity to explore the maximum mass allowed for
the stars.

\begin{table}[h]
  \centering
\begin{tabular}{ll}
  \hline
  $^A_Z X$	& $\rho\ ({\rm g\ cm^{-3}})$ \\
  \hline\hline
  $^4$He		 & $1.41\times 10^{11}$ \\
  $^{12}$C		 & $4.16\times 10^{10}$ \\
  $^{16}$O		 & $2.06\times 10^{10}$ \\
  $^{20}$Ne		 & $6.82\times 10^{9}$ \\
  $^{24}$Mg		 & $3.52\times 10^{9}$ \\
  $^{32}$S		 & $1.69\times 10^{8}$ \\
  $^{56}$Fe		 & $1.38\times 10^{9}$ \\
  \hline
\end{tabular}
\caption{Pressure and density threshold for which matter becomes
  unstable for four elements. The
  pressure values are taken from
  Ref. \cite{chamel/2015}. Considering the threshold pressure within the
  Hamada-Salpeter EoS, we have the corresponding density threshold.}
\label{limits}
\end{table}

\begin{figure}[h]
  \centering
  \begin{subfigure}{0.6\textwidth}
\includegraphics[scale=0.6]{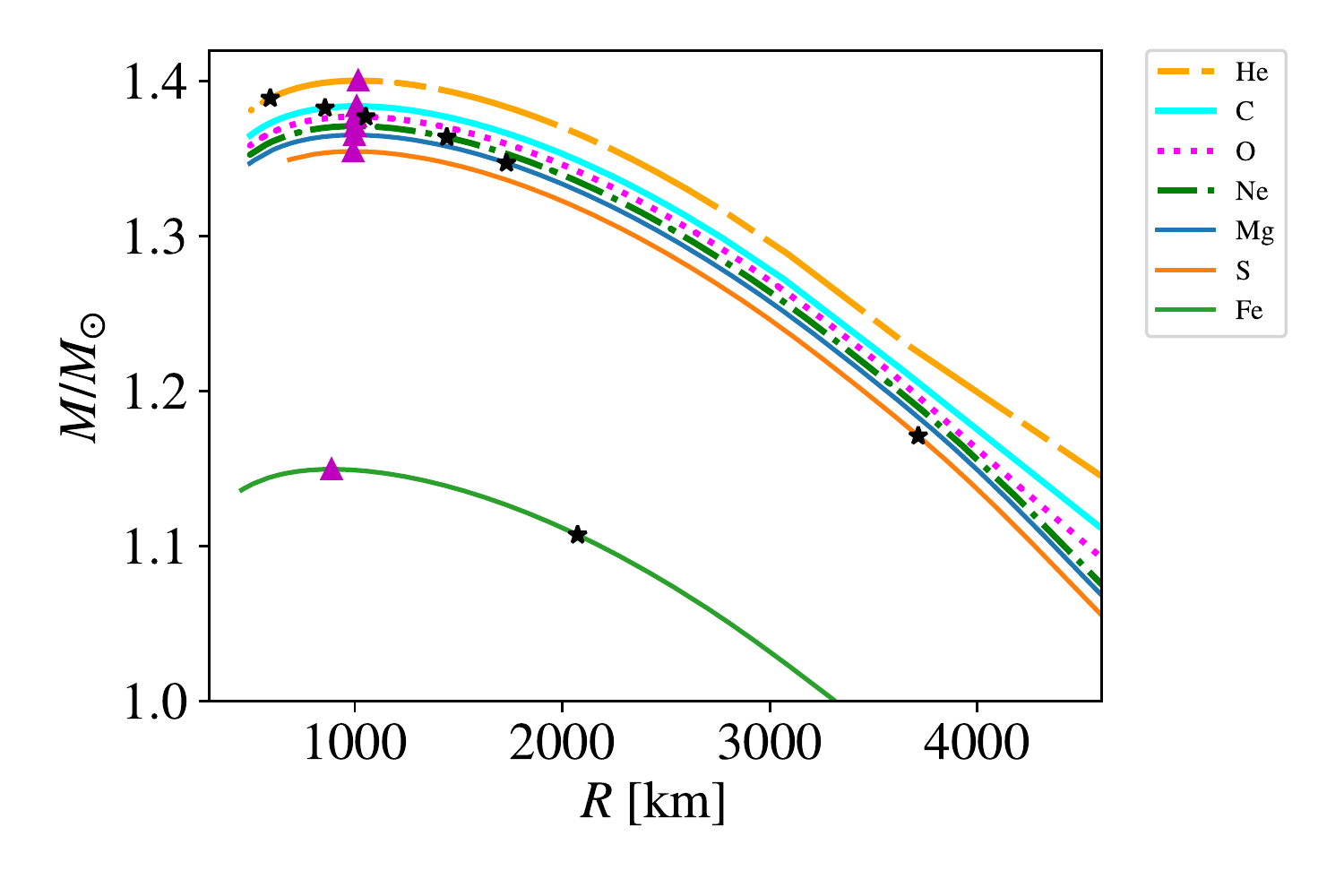}
\caption{Mass-radius relationship using the HS EoS for different star compositions.}
  \label{mrall}
\end{subfigure}
  \begin{subfigure}{0.6\textwidth}
  \includegraphics[scale=0.6]{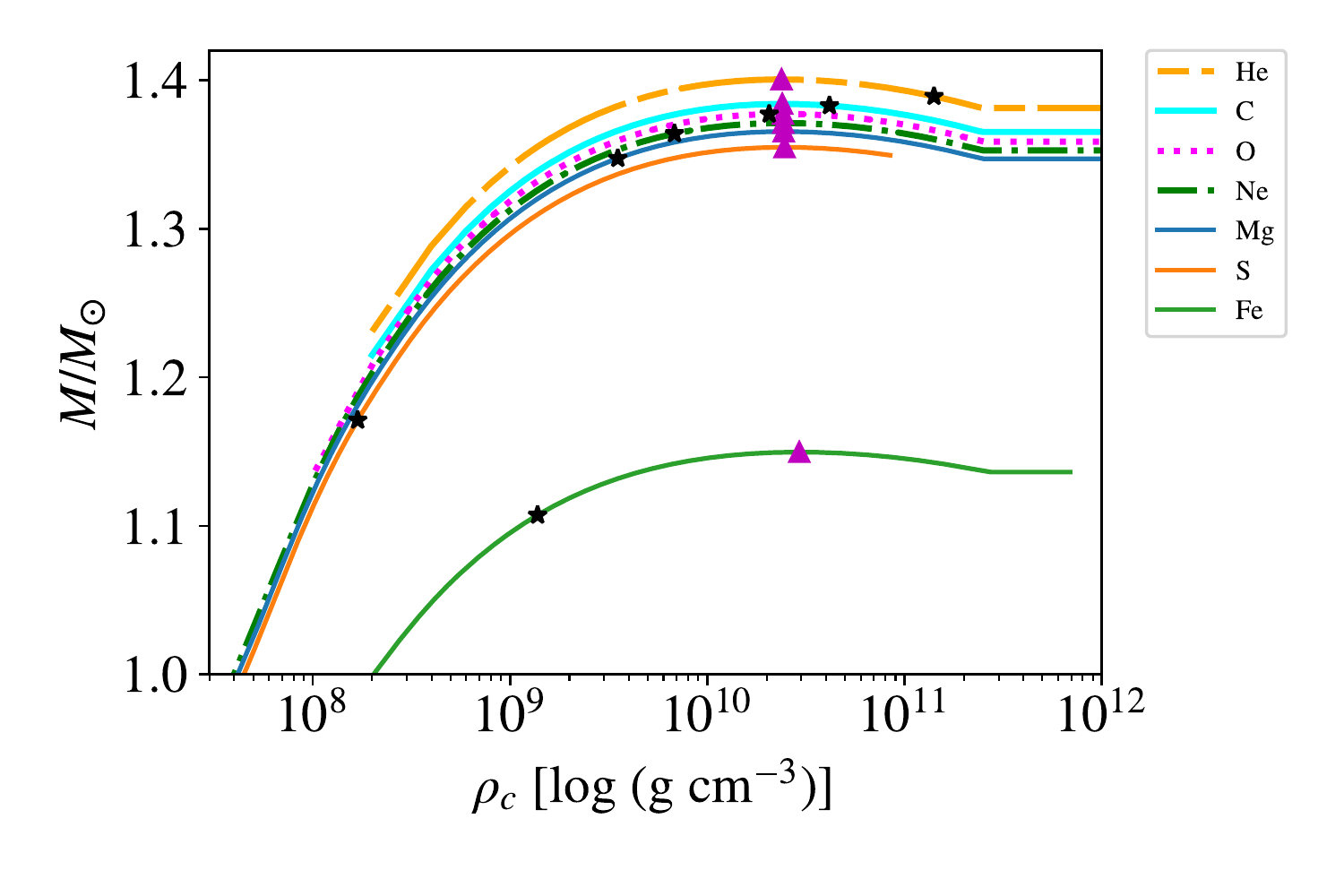}
  \caption{Mass vs central energy density using the Hamada-Salpeter EoS for different star compositions}
  \label{mrhoall}
\end{subfigure}
 \caption{White dwarfs with different EoS within general relativity.}
 \label{gr}
\end{figure}

In Fig. \ref{mrall}, we present the mass-radius
 for white dwarfs considering the HS EoS for different star's
 composition within general relativity. Pink triangles indicate the
 onset of the gravitational instability (maximum mass point) and black
 stars mark the onset of nuclear instabilities. The maximum mass is
 limited by the central density, therefore one needs to analyze it in a plot
 of mass-central energy density.

In Fig. \ref{mrhoall}, we show the behavior of the star's mass against
the central energy density within general relativity. Pink
triangles indicate the onset of the gravitational instability. From those points to the right, the stellar mass decreases with the
increment of $\rho_c$ and thus this region is unstable under radial oscillations.
Additionally, black stars mark the onset of nuclear instabilities.
From these points to the right side of the sequences, stars are
unstable due to electron capture reactions. As one can see, for light elements, the
gravitational instability limits the maximum mass of the star before the electron capture reactions can occur. However, for elements heavier than oxygen, the electron capture reactions take place before the maximum mass point is reached.
As a result, the nuclear instabilities are one of the main factors in restricting
the maximum stable mass.

\begin{figure}[h]
  \centering
  \begin{subfigure}{0.6\textwidth}
  \includegraphics[scale=0.6]{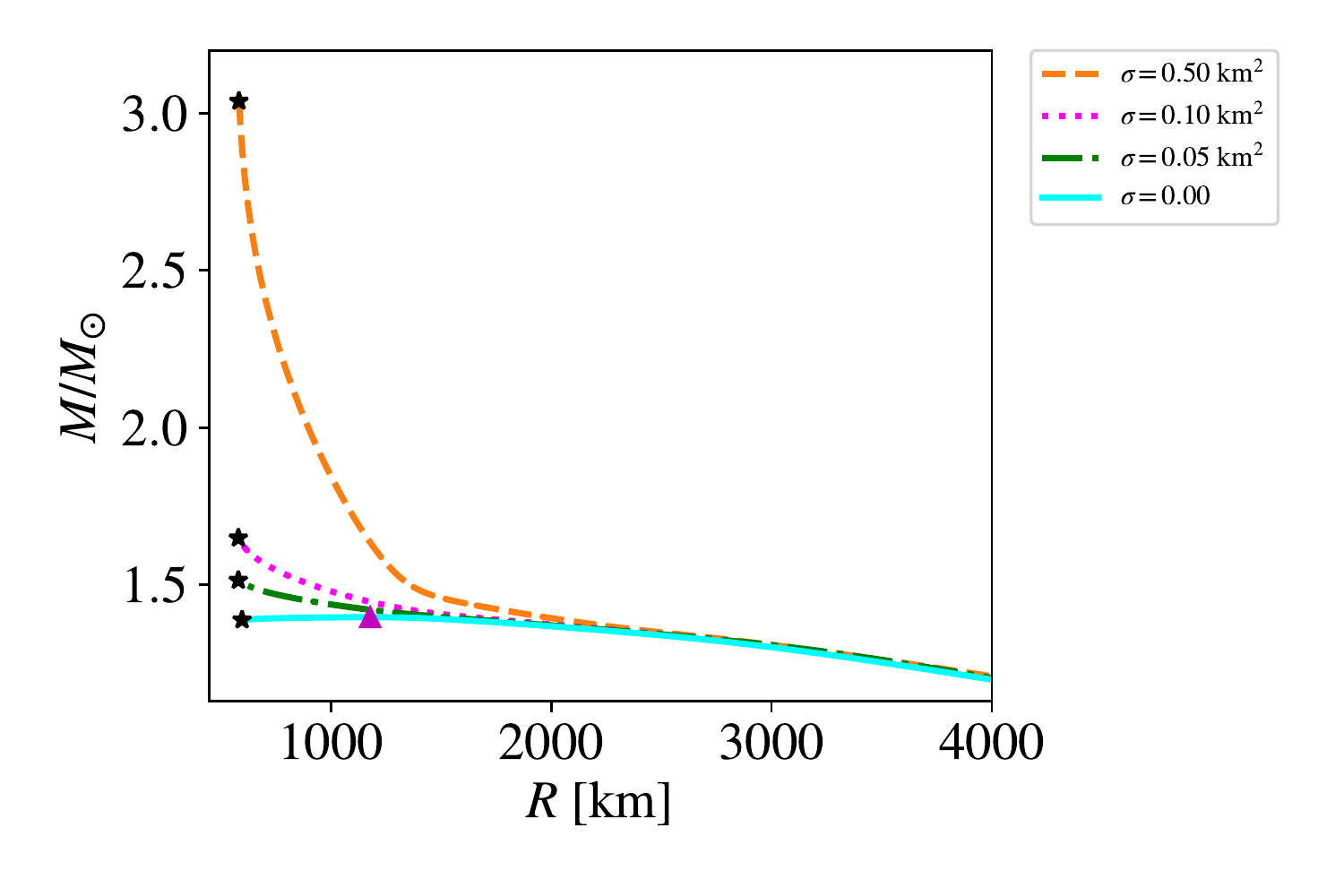}
  \caption{$^4$He WDs with
four different  values of modified gravity parameter.}
  \label{mr_frlhe}
\end{subfigure}
  \begin{subfigure}{0.6\textwidth}
    \includegraphics[scale=0.6]{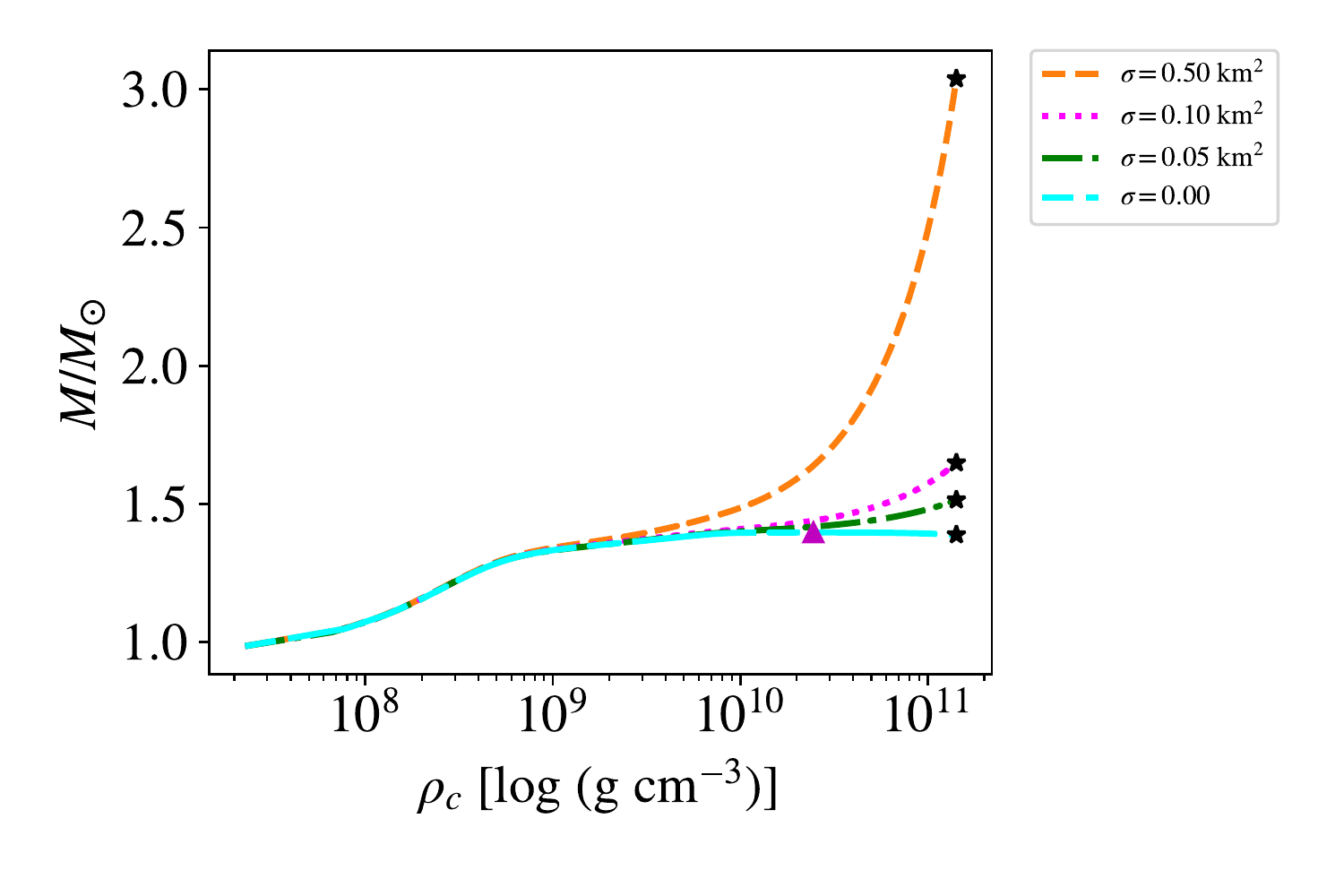}
  \caption{$^4$He WDs with
four different values of modified gravity parameter.}
  \label{rho_frlhe}
\end{subfigure}
 \caption{$^4$He white dwarfs within $\frl$ gravity.}
 \label{frl_he}
\end{figure}

In Fig. \ref{frl_he}, we show the mass-radius relationship and the sequence of stellar masses against the
central energy density within $\frl$ gravity for white dwarfs composed of
$^4$He. We have considered four
values for the theory's parameter. The values are: 0.00, 0.05, 0.10 and
0.50 ${\rm km^2}$. For $\sigma=0.00$ the theory recovers
the general relativity results.

In Fig. \ref{rho_frlhe}, we show the stellar masses against the
central energy density for the element $^4$He. We can see an increment in the masses according to the increase
in the value of $\sigma$. One can observe that when $\sigma\neq 0$ the stability criterion is not applicable and the gravitational instability disappears, i.e., the $dM/d\rho_c<0$ instability criterion is not achieved. Such a
behavior could imply in principle a white dwarf with an arbitrarily
large mass, which is an
unrealistic result given observational data. In this case, what constraints the maximum mass is the electron capture threshold marked by black stars.

\begin{figure}[h]
  \centering
  \begin{subfigure}{0.6\textwidth}
    \includegraphics[scale=0.6]{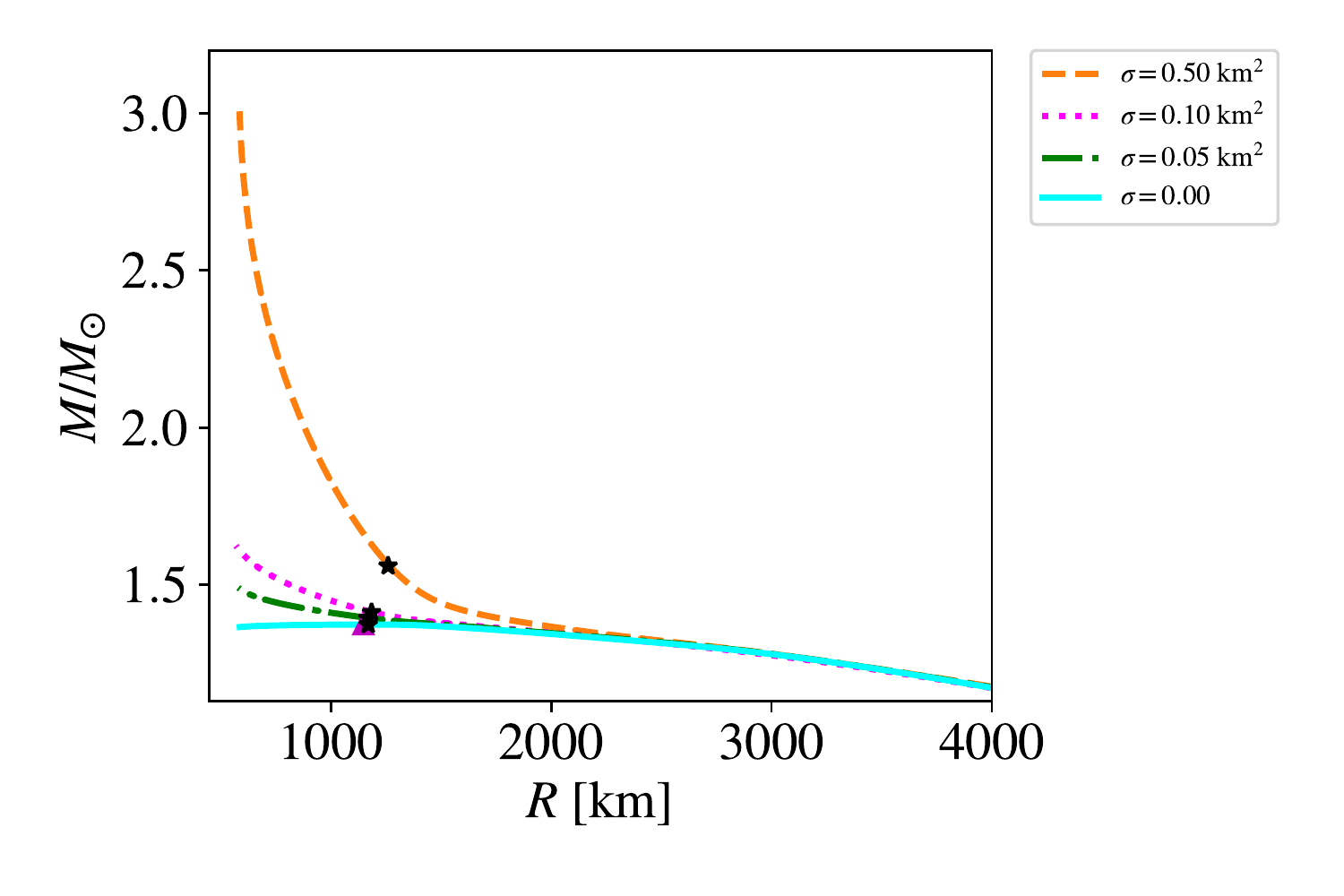}
  \caption{$^{16}$O WDs with
four different  values of modified gravity parameter.}
  \label{mr_frlO}
\end{subfigure}
  \begin{subfigure}{0.6\textwidth}
    \includegraphics[scale=0.6]{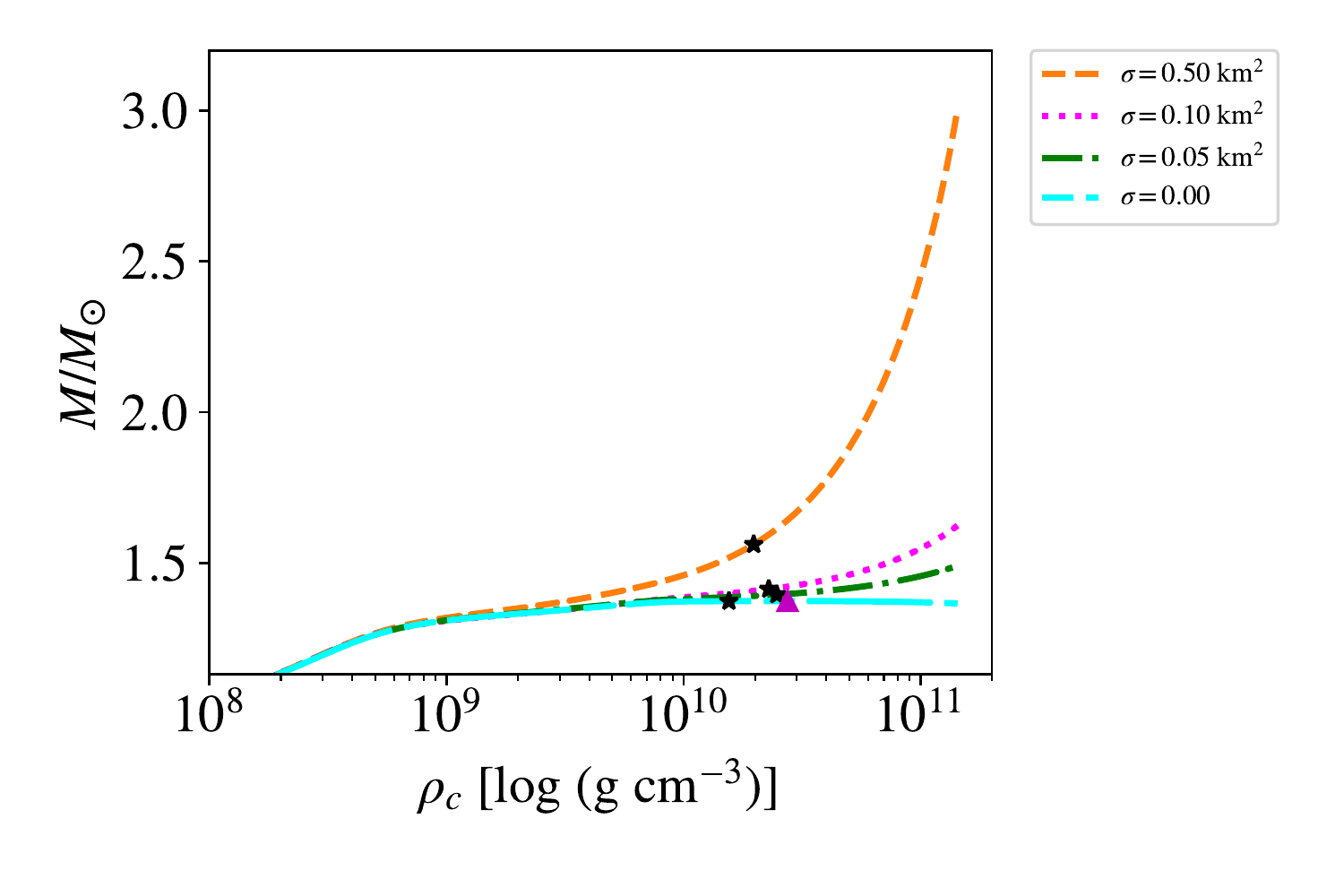}
  \caption{$^{16}$O WDs with
four different values modified gravity parameter.}
  \label{rho_frlO}
\end{subfigure}
 \caption{$^{16}$O white dwarfs within $\frl$ gravity.}
 \label{frl_O}
\end{figure}

In Fig. \ref{frl_O}, we show the mass-radius and mass-central density relations for $^{16}O$ WDs. For $^{16}O$, the density threshold for nuclear instabilities is smaller, which means that maximum stable mass is highly constrained by pycnonuclear reactions. In this case, we can see that the maximum becomes around 1.6 $M_\odot$. 

\begin{figure}[h]
  \centering
  \begin{subfigure}{0.6\textwidth}
    \includegraphics[scale=0.6]{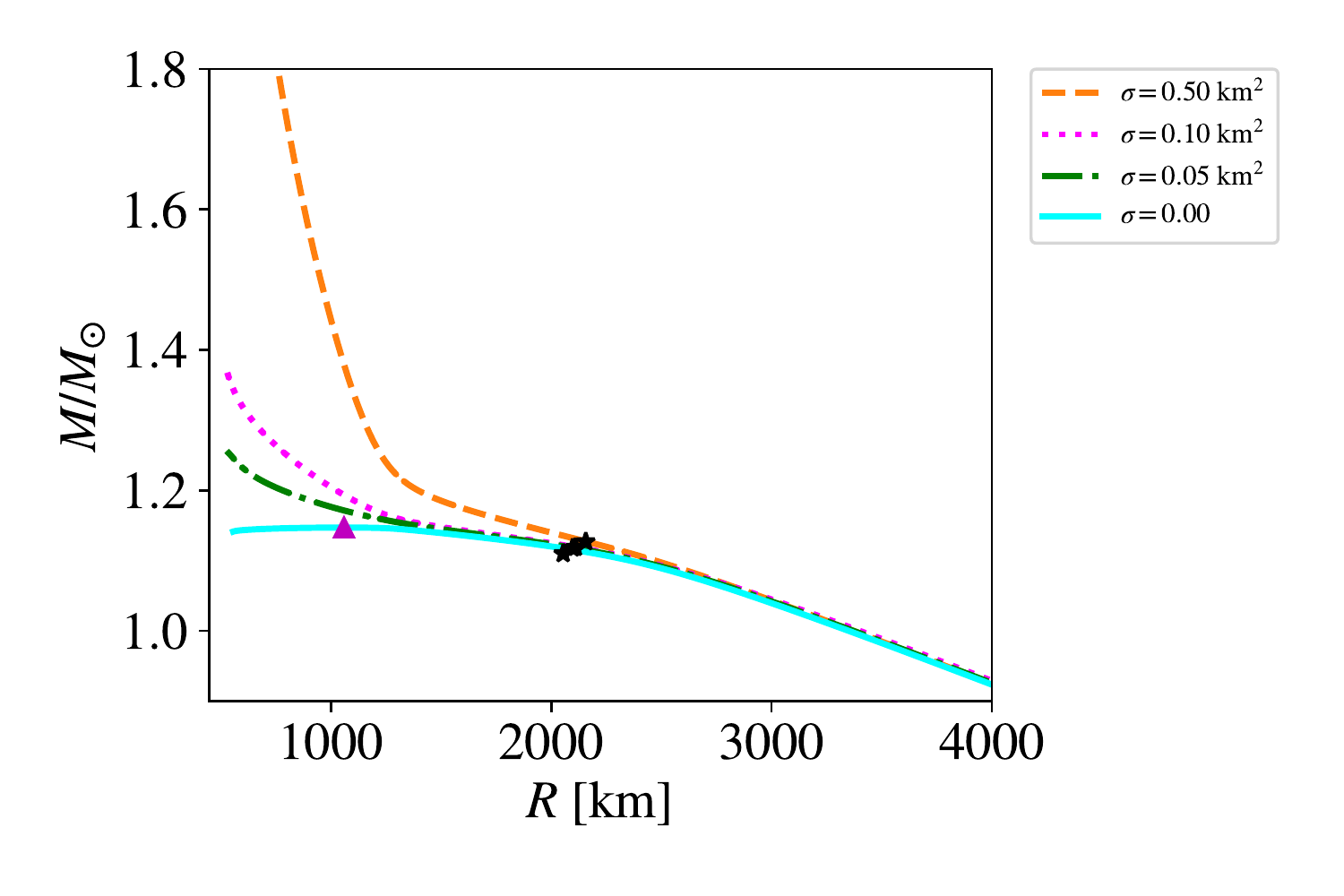}
  \caption{$^{56}$Fe WDs with
four different  values of modified gravity parameter.}
  \label{mr_frlFe}
\end{subfigure}
  \begin{subfigure}{0.6\textwidth}
    \includegraphics[scale=0.6]{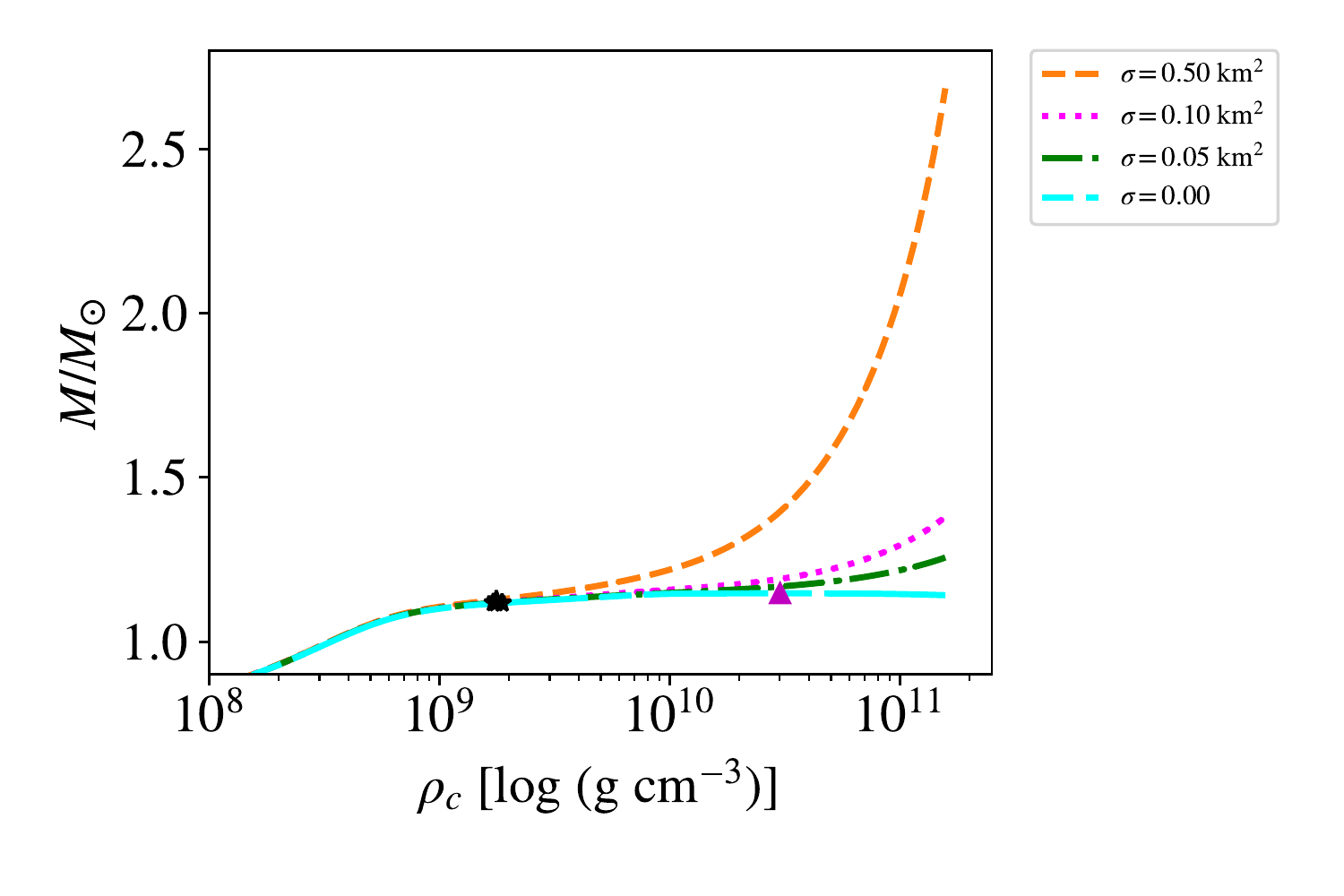}
  \caption{$^{56}$Fe WDs with
four different values modified gravity parameter.}
  \label{rho_frlFe}
\end{subfigure}
 \caption{$^{56}$Fe white dwarfs within $\frl$ gravity.}
 \label{frl_Fe}
\end{figure}

In Fig. \ref{rho_frlFe}, the element $^{56}$Fe was
considered in the stellar masses vs central energy density sequence. As in the previous case, increasing the theory's parameter also
leads to an enhancement in the maximum masses. However, as the density threshold for electron capture in $^{56}$Fe stars is remarkably smaller, the effects of the modified gravity theory for mass enhancement are almost negligible.

Hence, the density threshold for electron capture cannot be disregarded in dealing with modified theories of gravity and, in particular it drastically reduces the maximum stable mass. This is important in the context of modified theories of gravity being used to generate high stellar masses. Once there is a limit in the density regime, it must be respected or one will obtain misleading results.

\section{Concluding remarks}

  We have considered the stability of white dwarfs
within the $\mathtt{f(R,L_m)}$ theory of gravity. We found that
the standard gravitational Chandrasekhar limit is changed according to the increasing of the theory's parameter. However, instead of considering only gravitational instabilities for determining the maximum stable masses the nuclear instabilities must also be included, which leads to a remarkable decreasing of the maximum masses within modified theories.

\bibliographystyle{JHEP}
\bibliography{biblio}

\end{document}